\documentclass[english,preprint,aps,eqsecnum,superscriptaddress]{revtex4}
\usepackage[T1]{fontenc}
\usepackage[latin9]{inputenc}
\usepackage{xcolor}
\usepackage{pdfcolmk}
\usepackage{amsmath}
\usepackage{amssymb}
\usepackage{graphicx}
\PassOptionsToPackage{normalem}{ulem}
\usepackage{ulem}

\makeatletter

\providecommand{\tabularnewline}{\\}
\providecolor{lyxadded}{rgb}{0,0,1}
\providecolor{lyxdeleted}{rgb}{1,0,0}

\@ifundefined{textcolor}{}
{%
 \definecolor{BLACK}{gray}{0}
 \definecolor{WHITE}{gray}{1}
 \definecolor{RED}{rgb}{1,0,0}
 \definecolor{GREEN}{rgb}{0,1,0}
 \definecolor{BLUE}{rgb}{0,0,1}
 \definecolor{CYAN}{cmyk}{1,0,0,0}
 \definecolor{MAGENTA}{cmyk}{0,1,0,0}
 \definecolor{YELLOW}{cmyk}{0,0,1,0}
}

\usepackage{lscape}
\graphicspath{{Fig3/}}

\usepackage{babel}

\makeatother

\usepackage{babel}
\begin{document}

\title{Towards controlling the dissociation probability by light-induced
conical intersections}

\author{András Csehi }

\affiliation{Department of Theoretical Physics, University of Debrecen, H-4010
Debrecen, PO Box 5, Hungary}

\author{Gábor J. Halász }

\affiliation{Department of Information Technology, University of Debrecen, H-4010
Debrecen, PO Box 12, Hungary}

\author{Lorenz S. Cederbaum}

\affiliation{Theoretische Chemie, Physikalish-Chemisches Institut, Universität
Heidelberg, H-69120 Heidelberg, Germany}

\author{Ágnes Vibók }

\affiliation{Department of Theoretical Physics, University of Debrecen, H-4010
Debrecen, PO Box 5, Hungary}

\affiliation{ELI-ALPS, ELI-HU Non-Profit Ltd, H-6720 Szeged, Dugonics tér 13,
Hungary}
\begin{abstract}
Light-induced conical intersections (LICIs) can be formed both by
standing or by running laser waves. The position of a LICI is determined
by the laser frequency while the laser intensity controls the strength
of the nonadiabatic coupling. Recently, it was shown within the LICI
framework that linearly chirped laser pulses have an impact on the
dissociation dynamics of the $D_{2}^{+}$ molecule (J. Chem. Phys.
\textbf{143}, 014305, (2015); ibid \textbf{144}, 074309, (2016)).
In this work we exploit this finding and perform calculations using
chirped laser pulses in which the time dependence of the laser frequency
is designed so as to force the LICI to move together with the field-free
vibrational wave packet as much as possible. Since nonadiabaticity
is strongest in the vicinity of the conical intersection, this is
the first step towards controlling the dissociation process via the
LICI. Our showcase example is again the $D_{2}^{+}$ molecular ion.
To demonstrate the impact of the LICIs on the dynamical properties
of diatomics, the total dissociation probabilities and the population
of the different vibrational levels after the dissociation process
are studied and discussed.
\end{abstract}
\maketitle

\section{Introduction \vspace{0.5cm}
}

Conical intersections (CIs) provide a particularly striking and important
cause for nonadiabatic dynamics in polyatomic molecules not available
in diatomics. At a CI the nonadiabatic couplings become singular and
this gives rise to dramatic nonadiabatic effects widely studied in
the literature \cite{Graham1,Baer1}. Clearly, the position of a CI
appearing in nature and the strength of the nonadiabatic effects are
inherent properties of the electronic states of a molecule and are
hard to manipulate. For brevity, we rename the naturally occurring
CIs ``natural CIs''. While exposed to resonant laser light, a new
feature emerges. This feature is a CI induced by the light which cannot
be avoided even in the case of diatomic molecules \cite{Nimrod1,Milan1}.
The angle $\theta$ between the laser polarization and the molecular
axis now becomes the missing dynamical variable that together with
the stretching coordinate constitute the space in which the induced
CI can live. The phenomenon is rather general and not restricted to
propagating waves. Light-induced CIs (LICI) automatically emerge also
in standing waves which form optical lattices widely used in cold-atom
physics \cite{Nimrod1}. Actually, this is the case where they have
been first predicted. In optical lattices not only the electronic,
the stretching and rotational motion couple strongly via the induced
CI, but also the translational motion enters the scene and becomes
a dynamical variable. This gives cooling mechanisms new aspects. It
has been demonstrated that the LICI in diatomics gives rise to a variety
of nonadiabatic phenomena. Of course, they exhibit a topological phase
and provide singular nonadiabatic couplings like the natural CIs do
\cite{Gabi1,Gabi2}. Already in traveling waves of rather weak intensity
the LICI has a substantial impact on the spectrum of the molecule
\cite{Milan1}. Moreover, even in short laser pulses one finds remarkable
consequences of the LICI. The degree of alignment of a diatomic molecule
as a function of time differs substantially from the rigid rotor predictions
\cite{Gabi3}. A wealth of nuclear-wave-packet quantum interferences
show up in intense laser dissociation due to the nonadiabatic involvement
of the rotation \cite{Gabi4,Natan1}. A robust effect in the angular
distribution of the photofragments has been found that serves as a
direct signature of the LICI and should be amenable to observations
\cite{Gabi5}. If the resonant laser excitation is to an electronically
metastable state which decays by emitting an electron, the LICI is
in the continuum and changes its topology like all natural CIs which
are embedded in the continuum. It takes on the appearance of doubly-intersecting
complex surfaces (DICES) which also exhibit singular nonadiabatic
couplings \cite{Lenz1}. Examples where light-induced DICES appear
are resonant Auger decay in diatomics and interatomic Coulombic decay
by two-photon transitions in intense pulses. LICIs are ubiquitous
in polyatomic molecules both fixed in space or freely rotating. Because
of the presence of several vibrational degrees of freedom, LICIs exist
also without rotations and become multidimensional in nuclear coordinate
space \cite{Philip1}. This opens the door for manipulating and controlling
nonadiabatic effects by light \cite{Philip1,Kim,Ignacio1,Ignacio2,Albert,Kim1}. 

Currently, we are investigating the impact of linearly chirped laser
pulses on the dissociation of diatomic molecules (D$_{2}^{+}$) \cite{Andris1,Andris2}.
As the frequency of the pulse changes in time, the position of the
LICI changes analogously. It was found that using laser pulses which
are long enough ($T_{p}$$\geq$$10\, fs$) the amplitude of the periodic
change in the total dissociation probability is significantly squeezed
by the chirped pulse compared to the transform limited situation \cite{Andris1}.
By chirping the frequency the effective pulse duration is increased
and the effective intensity is decreased by a same factor resulting
in the suppression of the dissociation yields \cite{Andris2}. Several
works in the literature demonstrate the controllability of different
dynamical properties of molecular systems by using chirped laser pulses
\cite{Cao1,Cao2,krause,datta,Marangos,Adam1,Adam2,Adam3,kosloff,kosloff1,kosloff2,kosloff3,kosloff4,prob1,natan,prob2,chang,forre,zhang,Natan ,Noviitsky,Henriksen1,Henriksen2,Henriksen3}.

To be consistent with our former chirp type investigations, the showcase
example will again be the D$_{2}^{+}$ molecule. In the present work
we plan to go further and perform calculations using chirped laser
pulses in which the time dependence of the laser frequency is designed
so as to constrain for the motion of the LICI's position to follow,
as much as possible, the time evolution of the nuclear wave packet.
To achieve this goal, the time dependence of the laser frequency is
approximated by the energy difference between the potential energy
curves of the participating electronic states of the D$_{2}^{+}$
ion. We note, that the ideal case would be to let the LICI follow
the true wave packet and not the field-free one. This can be done
by a self-consistent procedure (optimal control theory). But, since
this is the first study of its kind, we would like to first see whether
there are gains to be expected from the LICI following the wave packet,
and this is the purpose the present investigation.

The photodissociation probabilities and the population of the different
vibrational levels after the dynamics are studied. Two dimensional
(2D) simulations are applied to the problem in which the rotational
angle is taken into account as a dynamic variable and, therefore,
the LICI is explicitly included.

\section{The Hamiltonian \vspace{0.5cm}
}

\begin{figure}[p]
\includegraphics[width=0.5\textwidth]{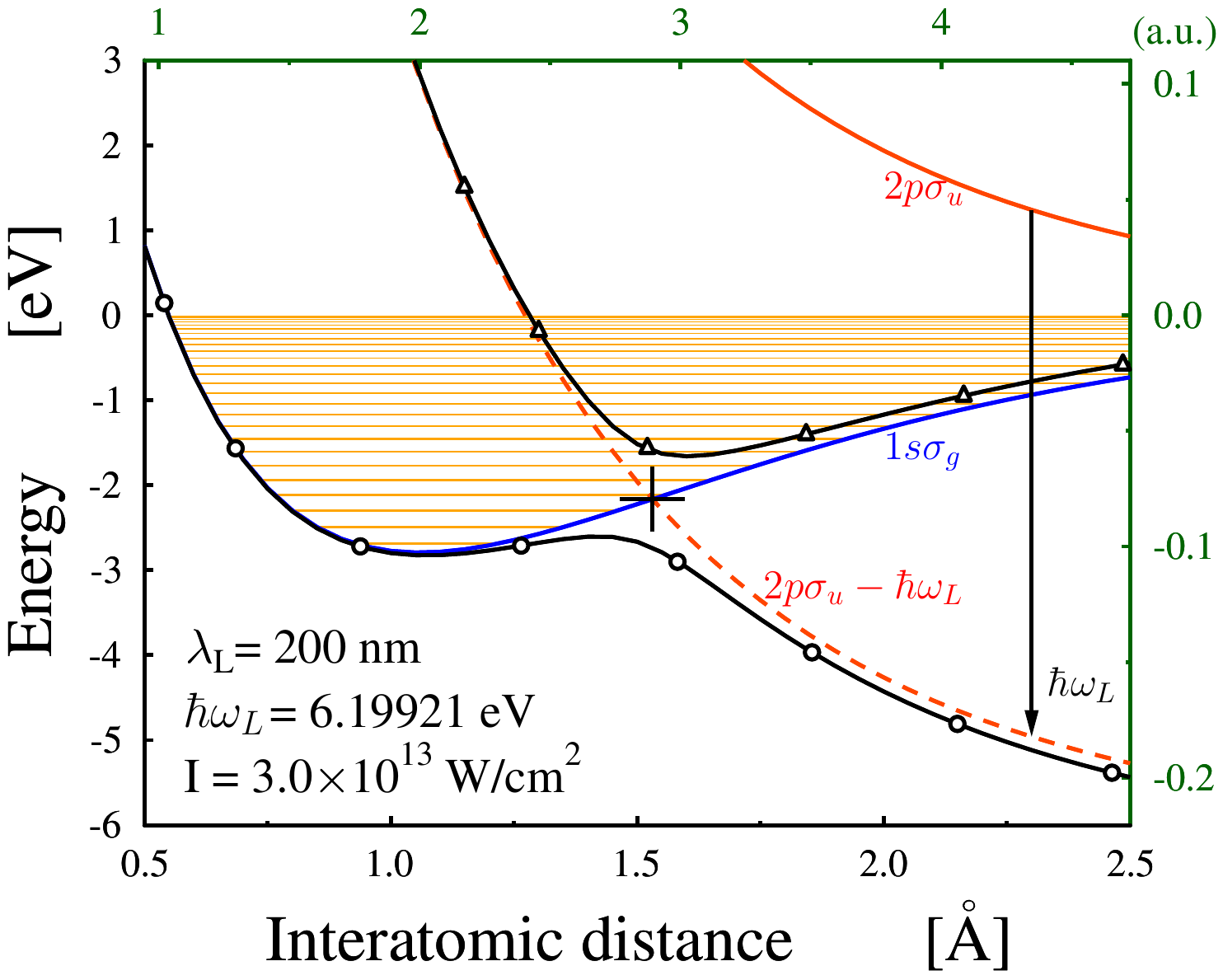}

\caption{\label{fig:potential}Potential energy curves of the D$_{2}^{+}$
ion. Diabatic energies of the ground $1s\sigma_{g}$ and the first
excited $2p\sigma_{u}$ states are displayed with solid blue and red
lines, respectively. The field dressed excited state ($2p\sigma_{u}$
- $\omega_{L}$ ) is presented with dashed red line. The field dressed
excited state ($2p\sigma_{u}$ - $\omega_{L}$ ) forms a light-induced
conical intersection (LICIs) with the ground state. A cut through
the adiabatic surfaces at $\theta=0$ (parallel to the field) is also
shown for a particular frequency ($\omega_{L}=6.19921\, eV$) and
field intensity ($3\times10^{13}\frac{W}{cm^{2}}$) depicted by solid
black lines marked with circles ($V_{l}$) and triangles ($V{}_{u}$).
Alternative x-y axises (top and right hand side) with a.u.'s are provided.
The horizontal orange lines denote the field-free vibrational levels
of the $1s\sigma_{g}$ state. We denote with a cross the position
of the LICI ($R_{LICI}=1.53\textrm{\AA}=2.891a.u.$ and $E_{LICI}=-2.16611eV$). }

\end{figure}
Figure \ref{fig:potential} shows the potential energy curves for
$D_{2}^{+}$ in Floquet representation \cite{Floquet}. $V_{1}(R)=1s\sigma_{g}$,
$V_{2}(R)=2p\sigma_{u}$ and ($V_{2}(R)-\hbar\omega_{L}$) denote
the energies of the electronic ground, the first excited and the dressed
eigenstates, which are considered in the calculations. Assuming the
scenario where the neutral $D_{2}$ molecule is ionized at $t=0$,
its vibrational ground state is transferred vertically to the potential
energy curve of the ground electronic state ($V_{1}=1s\sigma_{g}$)
of the ion. This defines the initial wave packet for the nuclear dynamics,
which can be considered as the Franck--Condon (FC) distribution of
all the vibrational states of the ion $D_{2}^{+}$. Exciting the $1s\sigma_{g}$
ground electronic state of the $D_{2}^{+}$ molecule by a resonant
laser pulse to the repulsive $2p\sigma_{u}$ state (see also Fig.
\ref{fig:potential}) these two electronic states are radiatively
coupled by the electric field. An electronic transition occurs due
to nonvanishing transition dipole moment and light-induced states
are formed. In the space of the $V_{1}(R)$ and $V_{2}(R)$ electronic
states the following time-dependent Hamiltonian can be written for
the rovibronic nuclear dynamics of the $D_{2}^{+}$: 

\begin{align}
H & =\left(\begin{array}{cc}
-\frac{1}{2\mu}\frac{\partial^{2}}{\partial R^{2}}+\frac{L_{\theta\varphi}^{2}}{2\mu R^{2}} & \;0\\
0 & \;-\frac{1}{2\mu}\frac{\partial^{2}}{\partial R^{2}}+\frac{L_{\theta\varphi}^{2}}{2\mu R^{2}}
\end{array}\right)+\label{eq:Hamilton}\\
 & \left(\begin{array}{cc}
V_{1}(R) & -E\left(t\right)d(R)\cos\theta\\
-E\left(t\right)d(R)\cos\theta & V_{2}(R)
\end{array}\right).\nonumber 
\end{align}
The off-diagonal elements of eq. (\ref{eq:Hamilton}) represent the
radiative couplings, where the electric field $E\left(t\right)$ characterizes
the laser field. $d(R)$ $\left(=-\left\langle \psi_{1}^{e}\left|\sum_{j}r_{j}\right|\psi_{2}^{e}\right\rangle \right)$
is the transition dipole matrix element and $\theta$ is the angle
between the polarization direction of the light and the direction
of the molecular axes. Here, R and ($\theta,\varphi$) are the molecular
vibrational and rotational coordinates, respectively, $\mu$ is the
reduced mass, and $L_{\theta\varphi}$ is the angular momentum operator
of the nuclei ($e=m_{e}=\hbar=1;$ atomic units are used throughout
the article). The potential energies $V_{1}(R)$ and $V_{2}(R)$ and
the transition dipole moment were taken from \cite{pot,tdm}. 

In the light-induced potential picture, after the absorption of one
photon, the energy of the $V_{2}(R)$ excited electronic state is
shifted downwards by $\hbar\omega_{L}$ -- where $\omega_{L}$ is
the laser frequency -- creating a crossing between the diabatic ground
and excited potential energy curves. After diagonalizing this diabatic
potential energy matrix (eq. (\ref{eq:Hamilton})) so called light-induced
adiabatic states $V_{lower}$ and $V_{upper}$ are formed which can
cross each other, creating a conical intersection whenever the conditions
$\cos\theta=0$, $(\theta=\pi/2)$ and $V_{1}(R)=V_{2}(R)-\hbar\omega_{L}$
are simultaneously met \cite{Nimrod1}. The position of this LICIs
is determined by the laser frequency while the laser intensity controls
the strength of the nonadiabatic coupling \cite{Nimrod1}. Increasing
the frequency moves the CI to a smaller internuclear distance and
to a lower energetic position while the opposite holds when decreasing
the frequency.

\section{The details of the calculations \vspace{0.5cm}
}

Here a short description is given of the time-dependent (TD) electric
fields of the pulses and of the wave packet propagation method used
to solve the TD nuclear Scrödinger equation employing the Hamiltonian
of eq. (\ref{eq:Hamilton}).

\subsection{The applied electric fields}

At time $t=0$ the system is ionized and the vibrational wave packet
on the ground ionic state starts to oscillate on the $V_{1}=1s\sigma_{g}$
potential curve. The initial nuclear wave packet is built up as the
FC distribution of the ionic vibrational states and the molecule was
initially in its rotational ground state (J=0 ). Then a laser pulse
is applied after a time delay of $t$$_{delay}$. A linear polarized
electric field is considered which is a product of a Gaussian envelope
and a cos($\phi(t)$) functions

\begin{equation}
E(t)=\epsilon_{0}f(t)\cos(\phi(t)).\label{eq:electric field}
\end{equation}
Here $\epsilon_{0}$ is the maximum amplitude of the electric field
and $f(t)=e^{-\frac{1}{2\sigma^{2}}(t-t_{delay})^{2}},(\sigma=T_{p}/\sqrt{4\log2})$)
is the Gaussian shape of the pulse. $T$$_{p}$ and $t$$_{delay}$
are the pulse duration (at FWHM) and the temporal maximum of the pulse,
respectively. $\omega(t)=\frac{d}{dt}\phi(t)$ is the laser frequency
and its time dependence is the control knob. The $\omega(t)=V_{2}(R)-V_{1}(R)$
function ($\hbar=1)$ controls the position of the nuclear coordinate
$R$ at which the resonance condition for the appearance of the LICI
is fulfilled. The initial nuclear wave packet is built up as the FC
distribution of the ionic vibrational states and its field-free time
evolution can be seen in Fig. \ref{fig:field-free}A. The time period
of the field-free oscillation is about $24\, fs$. As the wave packet
is a superposition of several different vibrational eigenstates, it
spreads during the time evolution relatively quickly, but exhibits
recurrences. In order to force the LICI to move together with the
field-free vibrational wave packet, we estimate the time-dependent
frequency $\omega(t)$ from the motion of the field-free wave packet
shown in Fig. \ref{fig:field-free}A.

In our calculations the pulse duration was chosen to be $T_{p}=6\, fs$
in all examples studied, $t_{delay}$ was varied in the range between
$0$ to $100\, fs$, and the phase of the electric field at time $t$
was derived according to the following formula:

\begin{equation}
\phi(t)=\int_{0}^{t}\omega(t')dt'.\label{eq:fi}
\end{equation}

In the present numerical study three kinds of chirped $\omega(t)$
functions were applied. They were obtained as explained in the following
and are shown in Fig. \ref{fig:field-free}B. The so called $\omega_{global}(t)$
function (red line on Fig. \ref{fig:field-free}B) is obtained by
following the global maximum of the density of the field-free wave
packet $\left(\omega_{global}(t)=V_{2}(R_{global}(t))-V_{1}(R_{global}(t))\right)$
shown in Fig. \ref{fig:field-free}A. This function is seen to exhibit
abrupt strong oscillations at certain times. In order to get rid of
these undesirable oscillations we applied a technique described below
and obtained the so-called $\omega_{local}(t)$ function (black line
in Fig. \ref{fig:field-free}B). Instead of following the global maximum
of the density of the nuclear wave packet, this function follows the
maximum value of the local density of the wave packet $\left(\omega_{local}(t)=V_{2}(R_{local}(t))-V_{1}(R_{local}(t))\right)$.
For deriving the $R_{local}(t)$ at time $t$, the maximum of the
product of the density $|\psi(R,t)|^{2}$ and an exponential weight
factor was considered rather than the maxima of the density themselves.
This was performed by maximizing the following expression at each
value of $t$:

\begin{equation}
|\psi(R,t)|^{2}e^{-\frac{1}{2\sigma_{R}^{2}}(R-R_{local}(t-\Delta t))^{2}}.\label{eq:exp}
\end{equation}
Here $R_{local}(t-\Delta t)$ is the position of maximum in the preceding
time step ($\Delta t=0.1$ fs) and $\sigma_{R}$ has a value of $0.77$
a.u. for $t<80$ fs and $0.28$ a.u. for later times. 

Finally, the third kind of chirped $\omega(t)$ function was simply
obtained by following the average value $<R>$ of the variable $R$
as a function of time using as usual $<R(t)>=<\psi(R,t)|R|\psi(R,t)>.$
This chirped frequency function is named $\omega_{<R>}(t)$. 

Since the pulses used in the present calculations are relatively short,
we also compare the results with those of pulses of fixed frequencies,
where for each time delay this constant frequency is chosen to correspond
to $\omega_{local}(t)$ at that time delay. We will address the respective
calculations as frequency varying transform limited (vTL) calculations.
In many experiments a particular pulse is used for all time delays,
and we have, therefore, also performed a reference transform limited
(TL) calculation where the central frequency is the same ($\omega$
= $0.23518\, a.u.$) at all time delays. 

In the simulations the intensities $I=1\times10^{12}\frac{W}{cm^{2}}$
and $I=1\times10^{13}\frac{W}{cm^{2}}$ were applied. We note that
the time integrals over the electric fields of the chirped pulses
applied in the present work are small but not exactly zero%
\footnote{For TL cases the ratio of the integrals relative to the Gaussian envelope
are below $10^{-8}$, while for the other frequency functions ($\omega_{local}(t)$
and $\omega_{global}(t)$) are below 1.5\%.%
}.

\subsection{The wave packet propagation }

The dissociation dynamics in the LICI framework is described by solving
the time-dependent nuclear Schrödinger equation (TDSE) with the Hamiltonian
$H$ described by eq. (\ref{eq:Hamilton}). For this the MCTDH (multi
configuration time-dependent Hartree) method has been utilized \cite{Dieter1,Dieter2,Dieter3,Dieter4,Dieter5}.
For characterizing the vibrational degree of freedom we have used
FFT-DVR (Fast Fourier Transformation-Discrete Variable Representation)
with $N_{R}$ basis elements distributed within the range from 0.1
a.u. to 80 a.u. for the internuclear separation. The rotational degree
of freedom was described by Legendre polynomials $\left\{ P_{J}(\cos\theta)\right\} _{j=0,1,2,\cdots,N_{\theta}}$.
These so called primitive basis sets ($\chi$) were used to represent
the single particle functions ($\phi$), which in turn were used to
represent the wave function: 
\begin{eqnarray}
\phi_{j_{q}}^{(q)}(q,t) & = & \sum_{l=1}^{N_{q}}c_{j_{q}l}^{(q)}(t)\;\chi_{l}^{(q)}(q)\qquad q=R,\,\theta\label{eq:MCTDH-wf}\\
\psi(R,\theta,t) & = & \sum_{j_{R}=1}^{n_{R}}\sum_{j_{\theta}=1}^{n_{\theta}}A_{j_{R},j_{\theta}}(t)\phi_{j_{R}}^{(R)}(R,t)\phi_{j_{\theta}}^{(\theta)}(\theta,t).\nonumber 
\end{eqnarray}
 The actual number of primitive basis functions in the numerical simulations
were chosen to be $N_{R}=2048$ and $N_{\theta}=61$ for the vibrational
and rotational degrees of freedom, respectively. On both diabatic
surfaces and for both degrees of freedom a set of $n_{R}=n_{\theta}=4,\cdots,20$
single particle functions were applied to form the nuclear wave packet
of the system. (The actual value of $N_{\theta}$ and $n_{R}=n_{\theta}$
was chosen depending on the peak field intensity $I_{0}$.) Attention
has been paid to the proper choice of basis so that convergence has
been reached in each propagation.

The TD nuclear wave packet is used to calculate the total dissociation
probability $P_{diss}$ and the populations. The probability reads

\begin{equation}
P_{diss}=\intop_{0}^{\infty}dt\left\langle \psi(t)\left|W\right|\psi(t)\right\rangle .\label{eq:dissociationprob}
\end{equation}
where $-iW$ is the complex absorbing potential (CAP) applied at the
last $10\, a.u$. of the vibrational ($R$) grid. 

To obtain the population of the individual vibrational levels we have
used two steps of the simulation. The first one is initiated from
the FC state averaged nuclear wave function and contains the interaction
to the external electric field and the dissociation process. It is
technically finished when the CAP eliminates the flying away portion
of the wave packet ($t=350\, fs$). The second step of the simulation
uses the final wave function of the first one as an initial wave function.
From this second simulation we calculate the autocorrelation function
as:
\begin{equation}
C(t)=<\psi(R,\theta,0)|\psi(R,\theta,T)>\label{eq:autocorr}
\end{equation}
where $T$ is the time period of the second simulation. Next, we calculate
the Fourier transform of this function to get the spectra of the bounded
wave function\cite{Schinke,Gabriel,Gatti}: 
\begin{equation}
\sigma(\omega)=\frac{1}{\pi}\int_{0}^{T}\Re\left[C(t)\cdot e^{i\omega t}\right]\cdot e^{-t/\tau}\cdot\cos^{2}\frac{\pi t}{2T}\cdot dt.\label{eq:szigma}
\end{equation}
The $\tau$ damping parameter was chosen to be $400\, fs$. Using
a long enough second simulation ($T=350\, fs$) we can resolve the
individual lines of the vibrational levels in the spectra. Finally,
we integrate these individual peaks to get the population of the corresponding
vibrational level:
\begin{equation}
P_{\nu}=\int_{(E_{\nu-1}+E_{\nu})/2}^{(E_{\nu}+E_{\nu+1})/2}\sigma(\omega)d\omega.\label{eq:population}
\end{equation}

\section{Results and discussion \vspace{0.5cm}
}

\begin{figure}[p]
\includegraphics[width=0.5\textwidth]{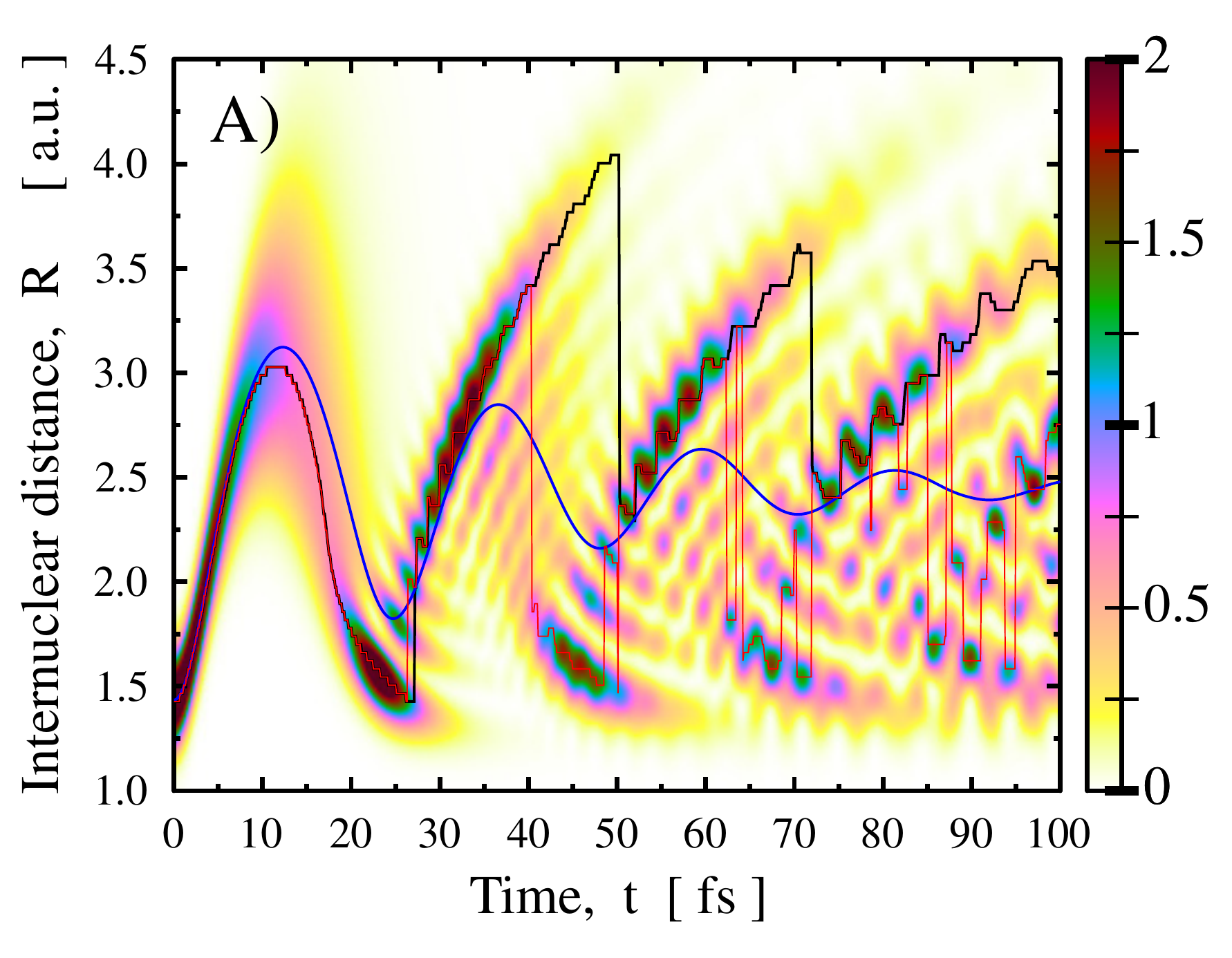}\includegraphics[width=0.5\textwidth]{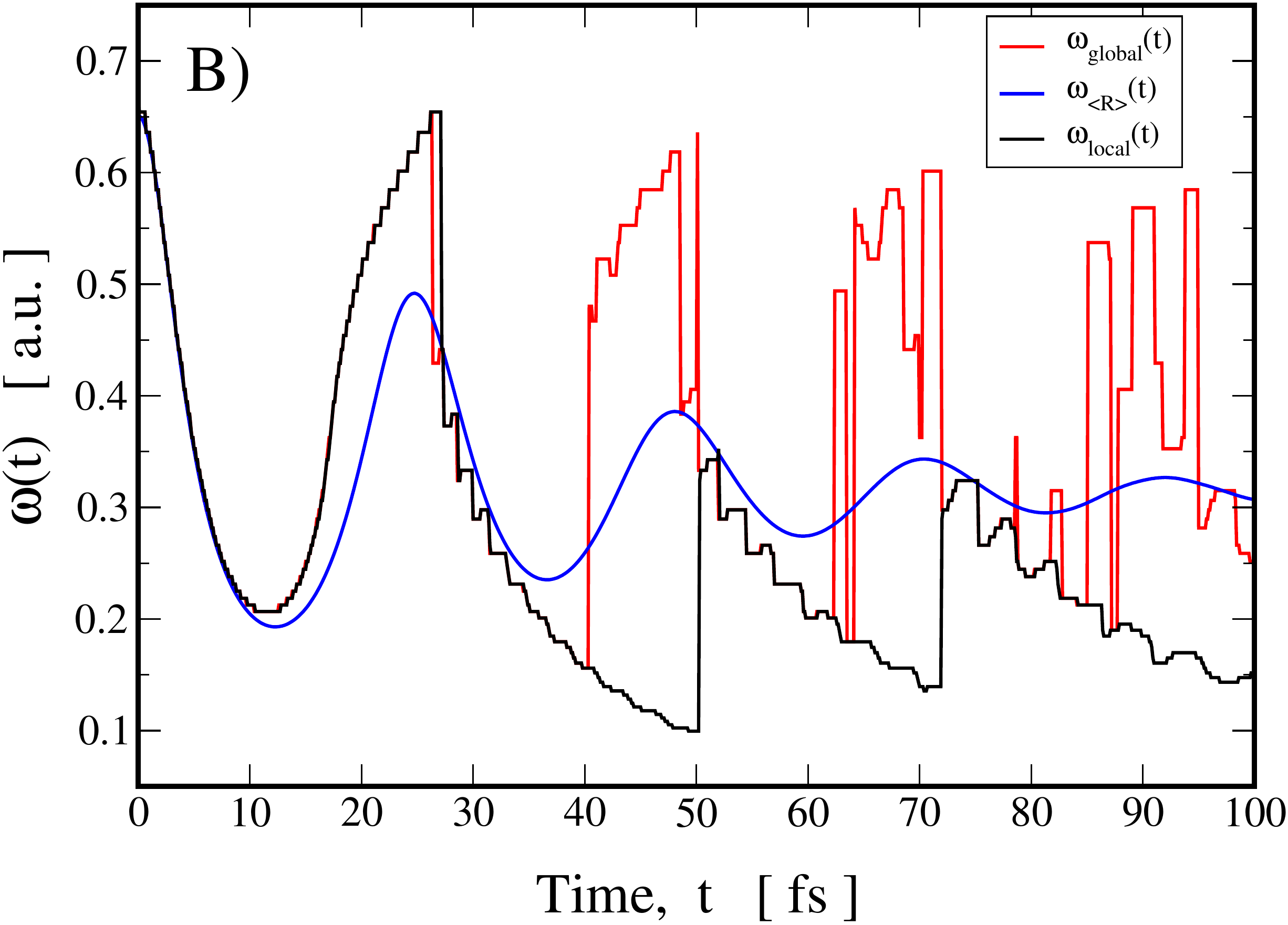}

\caption{\label{fig:field-free}The field-free vibration of the $D_{2}^{+}$
system as a function of time $t$ after ionization of the neutral
at time $t=0$ and the deduced chirped pulse frequency. In Panel (A)
the square of the vibrational wave packet is represented by color
code: the darker the color the larger the amplitude of the wave packet.
Additionally, three curves are shown which are used to determine the
chirped frequency of the pulses employed in this investigation: The
time-dependent average of the internuclear distance (blue curve),
a curve following the global (red curve) and one following the local
(black curve) maximum positions of the density of the field-free wave
packet. For more details see the main text. Panel (B) shows the resulting
three different forms of the $\omega(t)$ frequency function (for
definition see in Section IIIA.) of the laser pulse used in the calculations
of the nuclear dynamics. The respective frequency functions are denoted
by $\omega_{<R>}(t)\,,\omega_{global}(t)$ and $\omega_{local}(t)$. }
\end{figure}

In our previous work linearly chirped laser pulses were used, which
can easily be produced experimentally \cite{Pulse}. To be specific,
the pulses preserved the spectrum of their transform limited (TL)
equivalents and their time integral was exactly equal to zero. In
this work, our attention is on constructing pulses in which the time
dependence of the frequency is such that the light-induced conical
intersection moves together with the field-free vibrational wave packet.
This could be a first step towards developing a control protocol by
conical intersection, and it is meaningful, because in case of short
laser pulses the field-free wave packet does not change dramatically
during the dynamical process. We are interested in finding out how
much this can enhance the dissociation yield of the $D_{2}^{+}$ ion
and also studying how such a chirp influences the behavior of the
photodissociation probabilities as a function of delay time and the
populations on the different vibrational levels. Throughout the paper
$T_{p}=6\, fs$ pulse length is used.

\subsection{Dissociation probability}

\subsubsection{Chirped pulses}

\begin{figure}[p]
\includegraphics[width=0.5\textwidth]{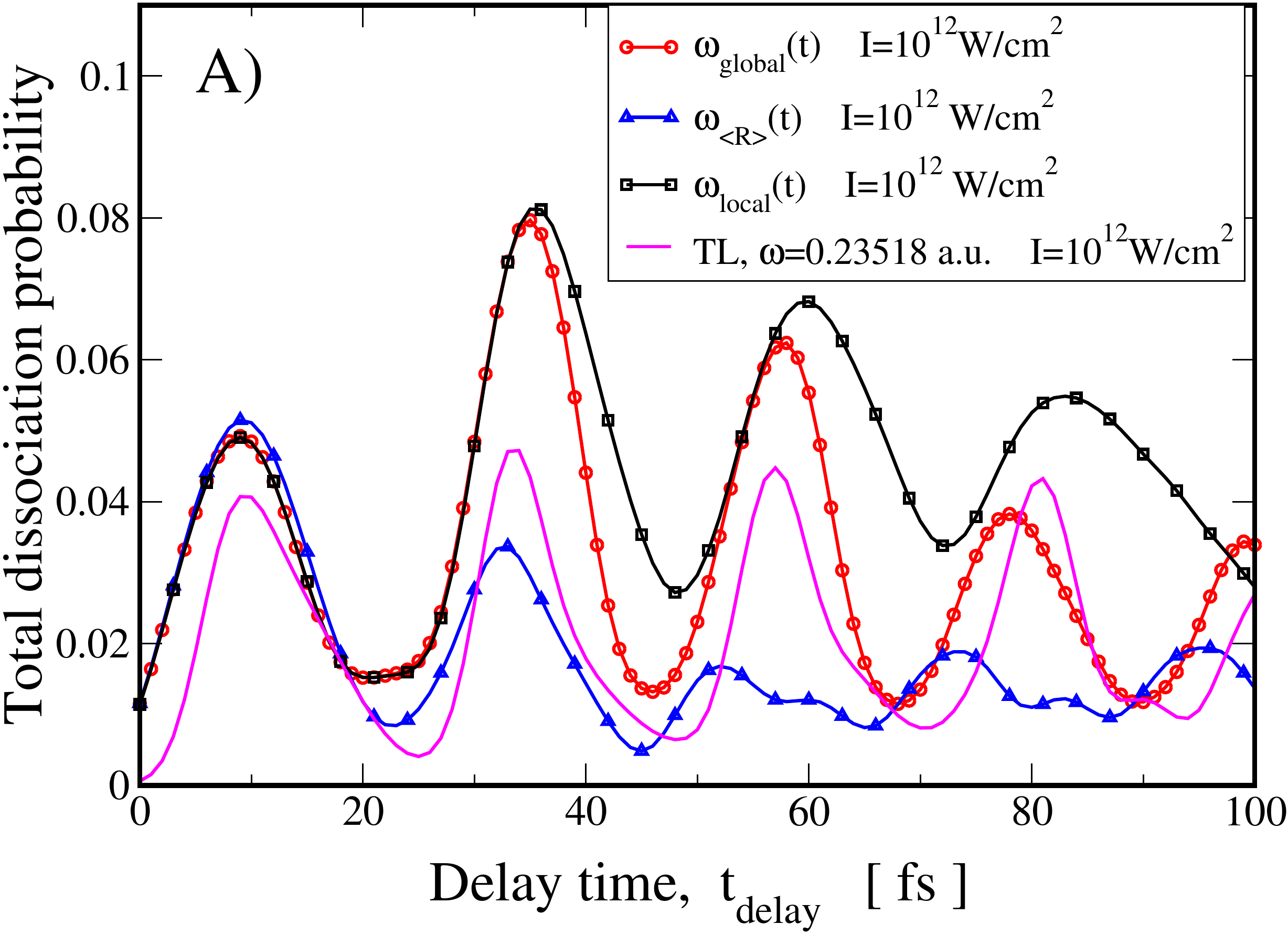}\includegraphics[width=0.5\textwidth]{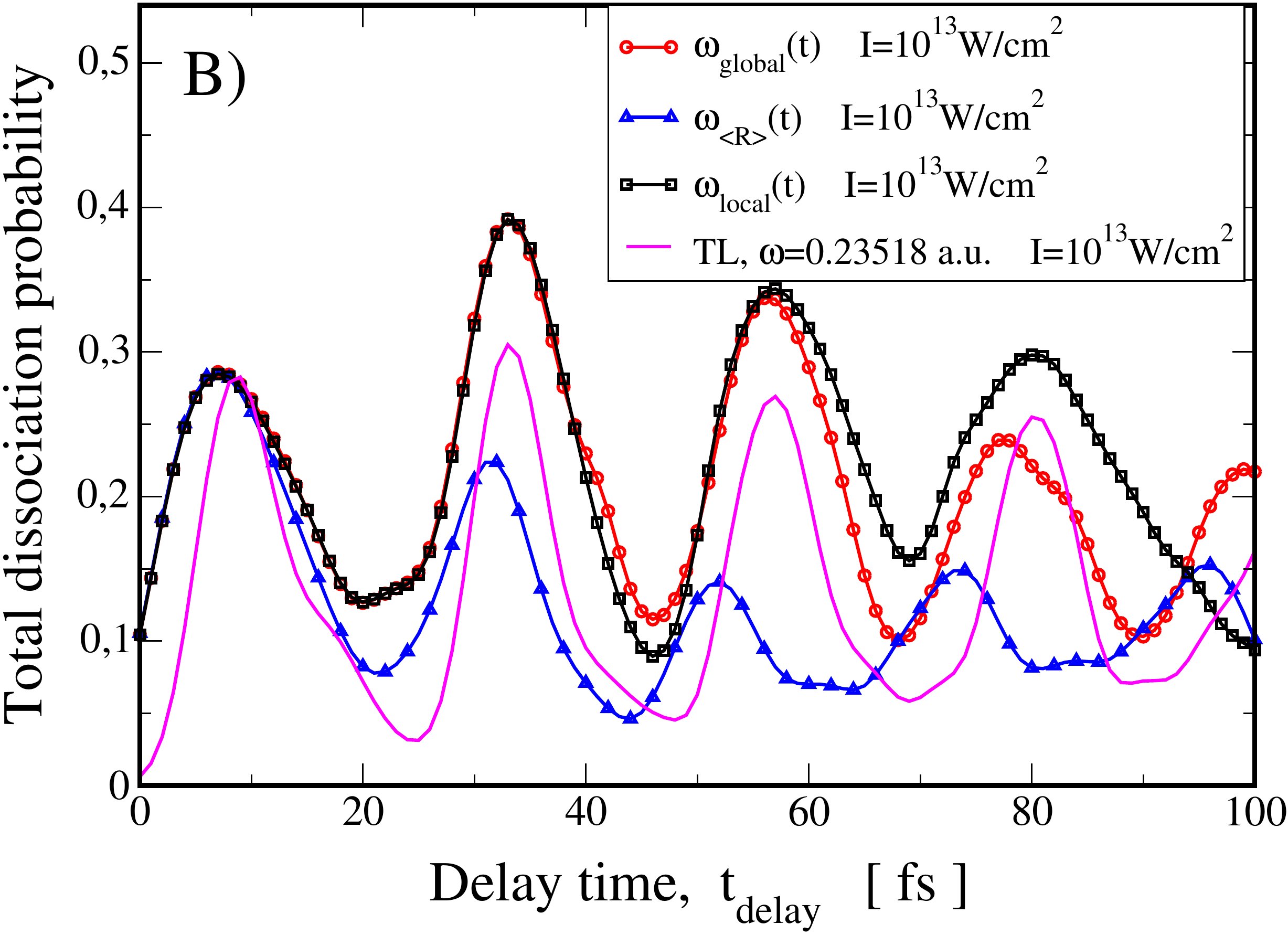}

\caption{\label{fig:3}Dissociation probabilities of D$_{2}^{+}$ as a function
of the delay time ($t_{delay}$) of the pulse. In panel (A) the black
(marked with square), the red (marked with circle) and the blue curves
(marked with triangle) denote the dissociation probabilities corresponding
to the $\omega_{local}(t),\,\omega_{global}(t)$ and $\omega_{<R>}(t)$
frequency functions, respectively, at $1\times10^{12}\frac{W}{cm^{2}}$
intensity. The fourth curve (magenta solid) shows the reference transform
limited (TL) result calculated with $\omega=0.23518\, a.u.$ Panel
(B) shows the same curves as in panel (A), but for $1\times10^{13}\frac{W}{cm^{2}}$
intensity. In all calculations the pulse length is $T_{p}=6\, fs$. }
\end{figure}

Figure \ref{fig:3} (panel A) shows the total dissociation probability
(Eq. (\ref{eq:dissociationprob})) as a function of time delay for
the TL and for the chirped simulations employing a $1\times10^{12}\frac{W}{cm^{2}}$
intensity. The central frequency of the TL pulse is the same for all
particular time delays and its value ($\omega$ = $0.23518\, a.u.$)
corresponds to the value of the second minimum of the $\omega_{<R>}(t)$
function in Fig. \ref{fig:field-free}B. Three different chirped $\omega(t)$
functions ($\omega_{global}(t)$, $\omega_{local}(t)$ and $\omega_{<R>}(t)$)
were applied in the simulations.

The most striking phenomenon is that each dissociation probability
curve shows a periodic oscillatory behavior. One can easily understand
it for the case of TL calculation, as in this situation the wave packet
is periodically located at a distance where the resonance condition
is satisfied, i.e. the LICI is positioned at the high density region
of the wave packet. One might expect more balanced results over the
time for the other curves, since in those cases the resonance condition
is almost always fulfilled in such places, where the density of the
wave packet possesses great value. Instead, we obtained curves which
are more or less in sync with the TL one, possessing well defined
maxima and minima and the ratios between these extreme values are
significantly large. The latter is true both for the $\omega_{global}(t)$
and $\omega_{local}(t)$ curves, but perhaps it is more relevant for
the case of $\omega_{global}(t)$. This is probably due to the finding
that at larger nuclear distances the resonance condition induces dissociation
more efficiently than at shorter distances. Indeed, starting, for
instance, from $t_{delay}=35\, fs$, when the position of the actually
selected local maximum of the nuclear density significantly differs
from the position of the global time maximum, the $\omega_{local}(t)$
curve gives the higher dissociation yield. Comparing Figures \ref{fig:field-free}A
and \ref{fig:3}A we see that the large values of the dissociation
rate correspond to large internuclear distances. That is, if the resonance
condition is satisfied at larger internuclear distances a more efficient
dissociation will result. Therefore, by determining the $\omega_{local}(t)$
function we have chosen that particular maximum value of the wave
packet density from all the possible maximum values which belongs
to a larger internuclear distance. It can be seen that the dissociation
rates obtained by the $\omega_{global}(t)$ and $\omega_{local}(t)$
functions go well together up to $32-34\, fs$ because in this case
the positions of the global and the chosen local maximum density values
coincide with each other. After $t_{delay}=34\, fs$, however, with
a few exceptions, the $\omega_{local}(t)$ function always provides
better results, i.e., higher dissociation probabilities. Although
the first strong deviation between the $\omega_{global}(t)$ and $\omega_{local}(t)$
functions occurs at around $t_{delay}=39\, fs$ , the impact of this
deviation on the dissociation rate has already been felt at the end
of the pulses centered at previous delay times. This manifests itself
in the dissociation yield from $t_{delay}=34\, fs$ onwards. 

Let us now discuss the results obtained with $\omega_{<R>}(t)$ frequency
function which has been determined by following the time-dependent
average of the internuclear distance. Although only marginally, the
dissociation probability obtained by the $\omega_{<R>}(t)$ frequency
function is the largest one up to three quarters of the first period
of the oscillation of the field-free nuclear wave packet ($t_{delay}=18\, fs$).
Accordingly, in this interval - when the wave packet is relatively
well localized - it is more beneficial to adjust the $\omega(t)$
function to the time-dependent average of the internuclear distance
$<R>$ than to the maximum of the nuclear wave packet density. For
$t_{delay}>18\, fs$ however, this dissociation yield (with $\omega_{<R>}(t)$)
is basically always the lowest and even by far so at large time delays.
 The reason for this behavior can be found in Figure \ref{fig:field-free}A.
It is clearly seen that for $t_{delay}>18\, fs$ the $\omega_{<R>}(t)$
function avoids the internuclear distances which correspond to the
largest global or any local nuclear density values. 

The dissociation probabilities for the pulses with $1\times10^{13}\frac{W}{cm^{2}}$
peak intensity are shown in Fig. \ref{fig:3}B. The situation is roughly
similar to that discussed above for the previous case of $1\times10^{12}\frac{W}{cm^{2}}$
peak intensity, but there are differences as well. It is eye-catching
that while the intensity has increased by a factor of 10, the maximum
yield of the dissociation increased only about 5-fold. This implies
that the dissociation probability has begun to saturate at this intensity.
 This is also related to the observation that while at the lower intensity
$\omega_{local}(t)$ provides the highest dissociation yields between
the $t_{delay}=35-50\, fs$, now at higher intensity, the yield curves
obtained with $\omega_{local}(t)$ and $\omega_{global}(t)$ go well
together up to $t_{delay}=40\, fs$. This saturation also affects
several other features of the curves. For example, while at lower
intensity in the $t_{delay}=40-50\, fs$ interval the $\omega_{local}(t)$
dissociation curve goes above the $\omega_{global}(t)$ one, the opposite
is true at the higher intensity. This is in spite of the finding that
the local nuclear density maxima are located at large internuclear
distances and the dissociation process is more efficient here. Due
to saturation, the large nuclear density present at small internuclear
distances together with the 10-fold higher intensity all together
produces more dissociation.  Saturation is also responsible for finding
that the differences between the dissociation yields obtained with
various frequency function are slightly smaller than found at low
intensity. All together, however, the use of $\omega_{local}(t)$
clearly provides the best choice for obtaining the largest dissociation
yield. 

In our previous studies \cite{Andris1,Andris2} the performance of
TL and linearly chirped laser pulses in obtaining the dissociation
probability has been investigated. There, the pulses used have fulfilled
several conditions. We stress again that in the present work the time-dependence
of the frequency functions of the pulses is designed in a completely
different way following a different objective. As a consequence, the
conditions put forward in \cite{Andris1,Andris2} do not hold anymore.
Due to the different choice of the chirped pulses used in this work,
a considerable growth of the total dissociation yield is obtained
compared to the former results (see Fig. 3B in \cite{Andris2}). 

\begin{figure}[p]
\includegraphics[width=0.5\textwidth]{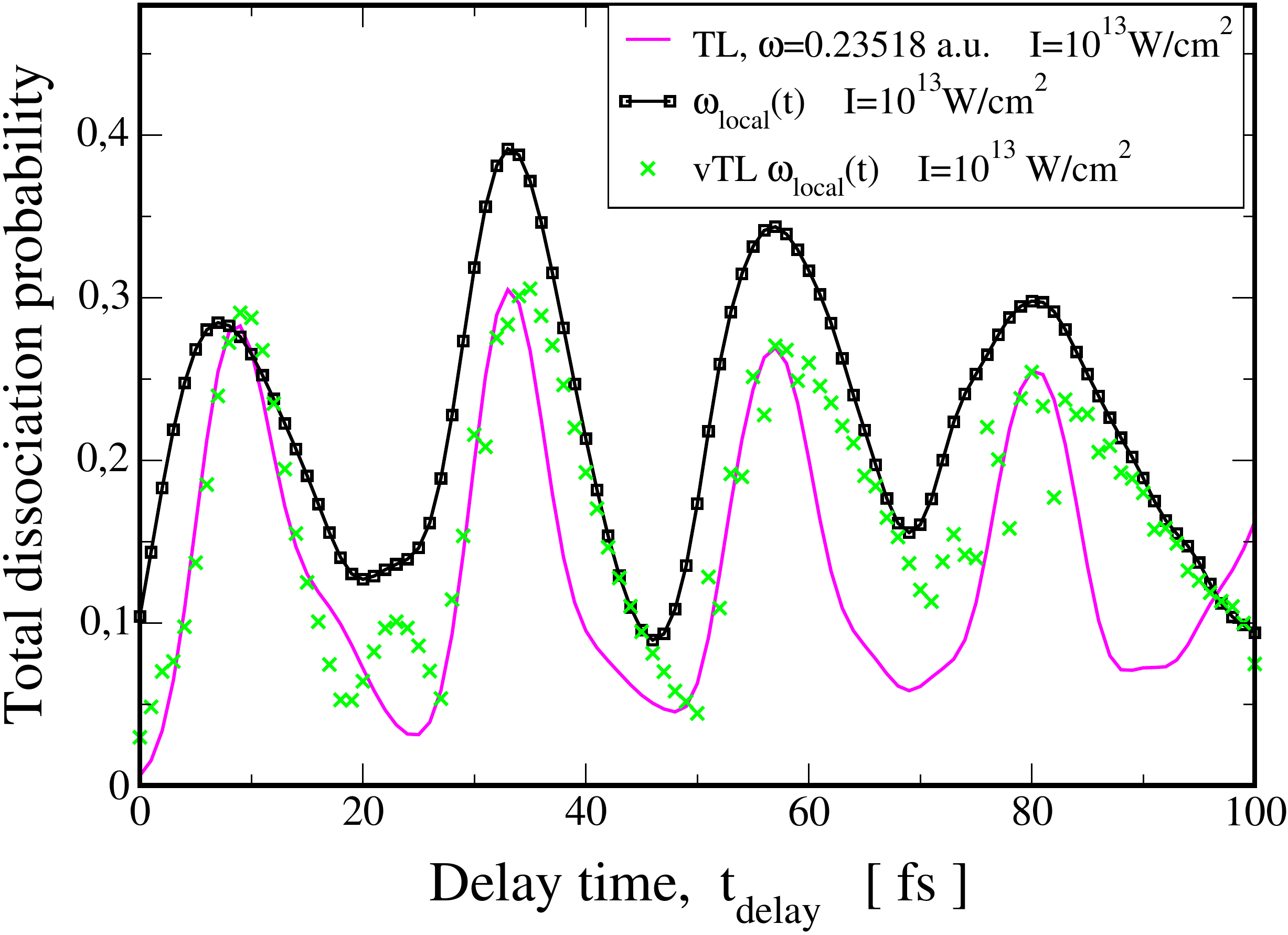}

\caption{\label{fig:4}Dissociation probabilities of D$_{2}^{+}$ as a function
of the delay time ($t_{delay}$) of the pulse. The black (marked with
square), the green crosses and the magenta (solid) curves show the
dissociation probabilities corresponding to the $\omega_{local}(t),$
vTL $\omega_{local}(t)$ and the TL $\omega=0.23518\, a.u.$ frequencies,
respectively. In the calculations the pulse length is $T_{p}=6\, fs$
and the intensity is $1\times10^{13}\frac{W}{cm^{2}}$.}
\end{figure}

\subsubsection{Varying transform limited pulses}

In the varying transform limited (vTL) calculations, the frequency
of the laser pulse was kept constant in every individual calculations
at each time delay, corresponding to the $\omega$ value given by
the $\omega_{local}(t)$ function at the respective time delay. The
results for the dissociation rates obtained by the vTL and $\omega_{local}(t)$
pulses as well as by the TL reference pulse are displayed in Fig.
\ref{fig:4}. 

Inspecting the curves, we see that across the whole time interval
the dissociation probabilities calculated by vTL pulses are smaller
than those obtained by $\omega_{local}(t)$, except for the tiny $t_{delay}=8-12\, fs$
interval. In this time period the LICI corresponding to the constant
frequencies of the vTL pulses is located at large internuclear distances
during the pulse duration providing a more efficient dissociation,
while in the case of the chirped pulse this applies only for the center
of the pulse. However, because the nuclear wave packet is well localized
in space during this time and its average momentum is close to 0,
there is no significant difference in the nuclear densities corresponding
to the two different positions of the LICI. This results in a higher
dissociation yield in this short time interval. 

Up to $t_{delay}=18\, fs$ the TL and vTL pulses provide very similar
results and this applies at many other time delays too. Such time
periods are e.g. $t_{delay}=28-30\, fs$ and $t_{delay}=50-55\, fs$
, at which the dissociation yield shows a growing trend. This is surprising,
because the vTL and TL pulses work on different frequencies. Mostly,
the dissociation yield obtained by the TL pulses exhibits a smooth
behavior. In particular beyond $t_{delay}=50\, fs$, the vTL results
do not lie on a smooth curve. This is related to the fact that several
islands of large density regions of the nuclear wave packet exist.
Since the vTL dissociation yield is calculated by varying the time
delay by $1\, fs$ steps, the frequency at two consecutive calculations
can refer to two different islands and thus be rather different leading
to the non-smooth behavior see in Fig. \ref{fig:4}. 

As time proceeds, the peaks in the dissociation yield as a function
of time delay become broader in the vTL calculations than in the reference
TL curve. This implies that it is beneficial to use a varying frequency
in order to increase the total dissociation probability. Although
the pulse duration is rather short ($T_{p}=6\, fs$), it is even more
beneficial to use the $\omega_{local}(t)$ frequency function where
the frequency changes during the pulse. As seen in Fig. \ref{fig:4},
the gain is then substantial.

\subsection{Populations of the individual vibrational levels after the pulse
is over}

\begin{table}[p]
{\small }%
\begin{tabular}{|c|c|c||c|c|c|c|c|c|}
\cline{4-9} 
\multicolumn{1}{c}{} & \multicolumn{1}{c}{} & \multicolumn{1}{c|}{} & \multicolumn{2}{c|}{{\small $t_{delay}=$8 fs}} & \multicolumn{2}{c|}{{\small $t_{delay}=$30 fs}} & \multicolumn{2}{c|}{{\small $t_{delay}=$59 fs}}\tabularnewline
\noalign{\vskip-1mm}
\hline 
{\small $\nu$ } & {\small $E_{\nu}$} & \multicolumn{1}{c|}{{\small $P_{\nu}^{FC}$}} & {\small $\Delta P_{\nu}$ } & {\small{} $\Delta P_{\nu}/P_{\nu}^{FC}$ } & {\small $\Delta P_{\nu}$ } & {\small $\Delta P_{\nu}/P_{\nu}^{FC}$ } & {\small $\Delta P_{\nu}$ } & {\small $\Delta P_{\nu}/P_{\nu}^{FC}$ }\tabularnewline
\hline 
\noalign{\vskip-1mm}
\hline 
\texttt{\small 0 } & \texttt{\small -2.69} & \texttt{\small 0.03655} & \texttt{\small -0.00009} & \texttt{\small ~-0.23\%} & \texttt{\small -0.00485} & \texttt{\small -13.27\%} & \texttt{\small ~0.00252} & \texttt{\small ~~6.90\%}\tabularnewline
\noalign{\vskip-2mm}
\hline 
\texttt{\small 1 } & \texttt{\small -2.49} & \texttt{\small 0.09619} & \texttt{\small -0.00489} & \texttt{\small ~-5.08\%} & \texttt{\small -0.02095} & \texttt{\small -21.78\%} & \texttt{\small ~0.00170} & \texttt{\small ~~1.77\%}\tabularnewline
\noalign{\vskip-2mm}
\hline 
\texttt{\small 2 } & \texttt{\small -2.30} & \texttt{\small 0.14372} & \texttt{\small -0.02038} & \texttt{\small -14.18\%} & \texttt{\small -0.04211} & \texttt{\small -29.30\%} & \texttt{\small -0.03123} & \texttt{\small -21.73\%}\tabularnewline
\noalign{\vskip-2mm}
\hline 
\texttt{\small 3 } & \texttt{\small -2.11} & \texttt{\small 0.16054} & \texttt{\small -0.04411} & \texttt{\small -27.48\%} & \texttt{\small -0.05510} & \texttt{\small -34.32\%} & \texttt{\small -0.06932} & \texttt{\small -43.18\%}\tabularnewline
\noalign{\vskip-2mm}
\hline 
\texttt{\small 4 } & \texttt{\small -1.94} & \texttt{\small 0.14976} & \texttt{\small -0.06280} & \texttt{\small -41.94\%} & \texttt{\small -0.05652} & \texttt{\small -37.74\%} & \texttt{\small -0.07537} & \texttt{\small -50.33\%}\tabularnewline
\noalign{\vskip-2mm}
\hline 
\texttt{\small 5 } & \texttt{\small -1.77} & \texttt{\small 0.12374} & \texttt{\small -0.06475} & \texttt{\small -52.32\%} & \texttt{\small -0.05176} & \texttt{\small -41.83\%} & \texttt{\small -0.05902} & \texttt{\small -47.69\%}\tabularnewline
\noalign{\vskip-2mm}
\hline 
\texttt{\small 6 } & \texttt{\small -1.61} & \texttt{\small 0.09385} & \texttt{\small -0.05115} & \texttt{\small -54.50\%} & \texttt{\small -0.04408} & \texttt{\small -46.97\%} & \texttt{\small -0.04995} & \texttt{\small -53.23\%}\tabularnewline
\noalign{\vskip-2mm}
\hline 
\texttt{\small 7 } & \texttt{\small -1.46} & \texttt{\small 0.06693} & \texttt{\small -0.03216} & \texttt{\small -48.05\%} & \texttt{\small -0.03289} & \texttt{\small -49.14\%} & \texttt{\small -0.04012} & \texttt{\small -59.94\%}\tabularnewline
\noalign{\vskip-2mm}
\hline 
\texttt{\small 8 } & \texttt{\small -1.31} & \texttt{\small 0.04564} & \texttt{\small -0.01569} & \texttt{\small -34.37\%} & \texttt{\small -0.01633} & \texttt{\small -35.78\%} & \texttt{\small -0.01018} & \texttt{\small -22.30\%}\tabularnewline
\noalign{\vskip-2mm}
\hline 
\texttt{\small 9 } & \texttt{\small -1.17} & \texttt{\small 0.03013} & \texttt{\small -0.00512} & \texttt{\small -16.98\%} & \texttt{\small -0.00154} & \texttt{\small ~-5.12\%} & \texttt{\small ~0.01627} & \texttt{\small ~54.01\%}\tabularnewline
\noalign{\vskip-2mm}
\hline 
\texttt{\small 10 } & \texttt{\small -1.04} & \texttt{\small 0.01945} & \texttt{\small ~0.00066} & \texttt{\small ~~3.37\%} & \texttt{\small ~0.00350} & \texttt{\small ~18.00\%} & \texttt{\small ~0.00371} & \texttt{\small ~19.10\%}\tabularnewline
\noalign{\vskip-2mm}
\hline 
\texttt{\small 11 } & \texttt{\small -0.92} & \texttt{\small 0.01237} & \texttt{\small ~0.00274} & \texttt{\small ~22.17\%} & \texttt{\small ~0.00071} & \texttt{\small ~5.76\%} & \texttt{\small -0.00420} & \texttt{\small -33.99\%}\tabularnewline
\noalign{\vskip-2mm}
\hline 
\texttt{\small 12 } & \texttt{\small -0.80} & \texttt{\small 0.00779} & \texttt{\small ~0.00309} & \texttt{\small ~39.71\%} & \texttt{\small -0.00055} & \texttt{\small ~-7.12\%} & \texttt{\small ~0.00352} & \texttt{\small ~45.20\%}\tabularnewline
\noalign{\vskip-2mm}
\hline 
\texttt{\small 13 } & \texttt{\small -0.70} & \texttt{\small 0.00489} & \texttt{\small ~0.00274} & \texttt{\small ~56.00\%} & \texttt{\small -0.00050} & \texttt{\small -10.31\%} & \texttt{\small -0.00033} & \texttt{\small ~-6.85\%}\tabularnewline
\noalign{\vskip-2mm}
\hline 
\texttt{\small 14 } & \texttt{\small -0.60} & \texttt{\small 0.00307} & \texttt{\small ~0.00209} & \texttt{\small ~68.17\%} & \texttt{\small -0.00027} & \texttt{\small ~-8.64\%} & \texttt{\small -0.00081} & \texttt{\small -26.24\%}\tabularnewline
\noalign{\vskip-2mm}
\hline 
\texttt{\small 15 } & \texttt{\small -0.50} & \texttt{\small 0.00193} & \texttt{\small ~0.00156} & \texttt{\small ~80.74\%} & \texttt{\small ~0.00012} & \texttt{\small ~~6.33\%} & \texttt{\small ~0.00057} & \texttt{\small ~29.37\%}\tabularnewline
\noalign{\vskip-2mm}
\hline 
\texttt{\small 16 } & \texttt{\small -0.42} & \texttt{\small 0.00123} & \texttt{\small ~0.00111} & \texttt{\small ~90.11\%} & \texttt{\small -0.00017} & \texttt{\small -13.86\%} & \texttt{\small ~0.00045} & \texttt{\small ~36.61\%}\tabularnewline
\noalign{\vskip-2mm}
\hline 
\texttt{\small 17 } & \texttt{\small -0.34} & \texttt{\small 0.00079} & \texttt{\small ~0.00077} & \texttt{\small ~98.13\%} & \texttt{\small ~0.00003} & \texttt{\small ~~4.36\%} & \texttt{\small ~0.00019} & \texttt{\small ~24.25\%}\tabularnewline
\noalign{\vskip-2mm}
\hline 
\texttt{\small 18 } & \texttt{\small -0.28} & \texttt{\small 0.00051} & \texttt{\small ~0.00053} & \texttt{\small 103.34\%} & \texttt{\small ~0.00002} & \texttt{\small ~~4.32\%} & \texttt{\small ~0.00005} & \texttt{\small ~10.12\%}\tabularnewline
\noalign{\vskip-2mm}
\hline 
\texttt{\small 19 } & \texttt{\small -0.21} & \texttt{\small 0.00033} & \texttt{\small ~0.00036} & \texttt{\small 106.16\%} & \texttt{\small -0.00009} & \texttt{\small -26.25\%} & \texttt{\small -0.00007} & \texttt{\small -22.66\%}\tabularnewline
\noalign{\vskip-2mm}
\hline 
\texttt{\small 20 } & \texttt{\small -0.16} & \texttt{\small 0.00022} & \texttt{\small ~0.00024} & \texttt{\small 109.22\%} & \texttt{\small ~0.00005} & \texttt{\small ~23.44\%} & \texttt{\small -0.00001} & \texttt{\small ~-4.16\%}\tabularnewline
\noalign{\vskip-2mm}
\hline 
\texttt{\small 21 } & \texttt{\small -0.11} & \texttt{\small 0.00015} & \texttt{\small ~0.00016} & \texttt{\small 108.50\%} & \texttt{\small ~0.00003} & \texttt{\small ~17.23\%} & \texttt{\small ~0.00005} & \texttt{\small ~33.79\%}\tabularnewline
\noalign{\vskip-2mm}
\hline 
\texttt{\small 22 } & \texttt{\small -0.08} & \texttt{\small 0.00010} & \texttt{\small ~0.00011} & \texttt{\small 106.53\%} & \texttt{\small -0.00003} & \texttt{\small -28.77\%} & \texttt{\small -0.00002} & \texttt{\small -21.05\%}\tabularnewline
\noalign{\vskip-2mm}
\hline 
\texttt{\small 23 } & \texttt{\small -0.05} & \texttt{\small 0.00006} & \texttt{\small ~0.00007} & \texttt{\small 104.77\%} & \texttt{\small -0.00002} & \texttt{\small -33.46\%} & \texttt{\small ~0.00001} & \texttt{\small ~10.44\%}\tabularnewline
\noalign{\vskip-2mm}
\hline 
\texttt{\small 24 } & \texttt{\small -0.02} & \texttt{\small 0.00004} & \texttt{\small ~0.00004} & \texttt{\small 100.79\%} & \texttt{\small ~0.00000} & \texttt{\small ~-7.41\%} & \texttt{\small ~0.00002} & \texttt{\small ~59.29\%}\tabularnewline
\hline 
\noalign{\vskip-2mm}
\multicolumn{1}{|c|}{\texttt{\small 25 }} & \texttt{\small -0.01} & \texttt{\small 0.00003} & \texttt{\small ~0.00003} & \texttt{\small ~97.74\%} & \texttt{\small ~0.00000} & \texttt{\small ~14.39\%} & \texttt{\small ~0.00000} & \texttt{\small ~13.71\%}\tabularnewline
\hline 
\noalign{\vskip-2mm}
\multicolumn{3}{c}{\texttt{\small $\sum_{\nu}\Delta P_{\nu}$ :}} & \multicolumn{1}{c}{\texttt{\small -0.28483 }} & \multicolumn{1}{c}{\texttt{\small{} }} & \multicolumn{1}{c}{\texttt{\small -0.32329 }} & \multicolumn{1}{c}{\texttt{\small{} }} & \multicolumn{1}{c}{\texttt{\small -0.31155 }} & \multicolumn{1}{c}{}\tabularnewline
\hline 
\noalign{\vskip-2mm}
\multicolumn{3}{c}{{\small $E_{CI}(t=t_{delay})$ {[}eV{]} :}} & \multicolumn{1}{c}{\texttt{\small -2.26379 }} & \multicolumn{1}{c}{} & \multicolumn{1}{c}{\texttt{\small -2.48587 }} & \multicolumn{1}{c}{} & \multicolumn{1}{c}{\texttt{\small -2.18610 }} & \multicolumn{1}{c}{}\tabularnewline
\hline 
\noalign{\vskip-2mm}
\end{tabular}{\small \par}

\caption{The changes of the populations of the individual vibrational levels
during the light matter interaction for three different simulations.
In all three cases the intensity of the field was $1\times10^{13}W/cm^{2}$
and the frequency of the electric field was applied according to the
$\omega_{local}(t)$ function. In the first three columns the vibrational
levels ($\nu$), the corresponding energies ($E_{\nu}$) and the initial
populations ($P_{\nu}^{FC}$) of the vibrational eigenstates are displayed.
The next three main columns each containing two sub-columns contain
the absolute and relative changes of the populations for the chosen
$t_{delay}=8,\,30$ and $59\, fs$. In the bottom of the table the
last two rows contain the sum of the absolute changes (the dissociation
probability) and the energetic position of the LICI at time $t=t_{delay}$. }
\end{table}

We now examine the effect of photodissociation dynamics initiated
by the $\omega_{global}(t)$ chirped laser pulses for the population
of the different individual vibrational levels of the residual $D_{2}^{+}$
ion which has not undergone dissociation as the pulse terminated.
Comparison has been made between the initial FC population of the
$D_{2}^{+}$ molecule and the population of the different vibrational
levels after the dissociation has taken place. The results are collected
in Table 1. 

At the bottom of the table the actual energy positions of the LICI
corresponding to the relevant delay times are indicated. At $t_{delay}=8\, fs$
and $t_{delay}=59\, fs$ the LICI lies energetically between the $\nu=2$
and $\nu=3$ vibrational levels, while at $t_{delay}=30\, fs$ it
lies very close to the $\nu=1$ level. It is striking that in the
close vicinity of the LICI the population drastically decreases and
the depletion is continuously held up to the $\nu=8$ vibrational
level. This is the interval from where the ion can easily dissociate.
Then, above the $\nu=8$ vibrational level a sudden change occurs
resulting in a different behavior of the populations for the studied
delay times. While for $t_{delay}=8\, fs$ the population is continuously
increasing above the $\nu=9$ vibrational level, for the other two
delay times an alternating increase or decrease of the populations
occur. Because of the strong nonadiabatic effects present it is difficult
to explain this interesting reshuffling of the populations. Probably,
the presence of the nonadiabatic effects can, in addition to the direct
dissociation process, induce vibrational excitations resulting in
a newly mixed population distribution among the various vibrational
levels. This process can be dominant above $\nu=8$ level because
in this region practically no dissociation occurs. The populations
of the vibrational levels should be amenable to measurements.

\section{Conclusions\vspace{0.5cm}
}

Applying the light-induced conical intersection framework, numerical
simulations have been performed for studying the photodissociation
dynamics of the $D_{2}^{+}$ molecular ion in order to understand
better the relation between the evolving nuclear vibrational wave
packet and the frequency chirping of the laser pulse. We presented
a strategy to directly follow the time evolution of the nuclear wave
packet meanwhile forcing it to move together with the LICI as much
as possible. Based on this idea three different types of chirped frequencies
were designed. The effects of the laser pulses constructed by using
these $\omega_{global}(t)$, $\omega_{local}(t)$ and $\omega_{<R>}(t)$
functions have been explored in terms of the total dissociation rate
as a function of the time delay between the ionization of D$_{2}$
and the pulse, as well as in terms of the population of the different
vibrational levels after the pulse has terminated. We have found that
of all $\omega$ functions applied so far, the $\omega_{local}(t)$
provides the overall highest yield for the dissociation rate, which
is probably close to the largest possible outcome or dissociation
limit for the given pulse length and intensity values.

It is found that employing a frequency which is adapted to the time
delay in question is very beneficial for the dissociation rate. Moreover,
although the pulses are rather short ($T_{p}=6\, fs$), it is clearly
demonstrated that varying the frequency during the pulse, i.e., using
a chirped pulse, gives rise to a further large gain in the dissociation
rate. The reshuffling of the populations of the individual vibrational
levels of the residual pulse after the pulse has terminated has been
found to depend markedly on the time delay employed. These populations
contain much information on nonadiabatic effects due to the LICI and
should be amenable to experiments.

In our work the motion of the LICI was adjusted to the time evolution
of the nuclear field-free wave packet. As we applied only relatively
short laser pulses ($T_{p}=6\, fs$) it seems meaningful as a first
approximation to use the field-free wave packet because the pulses
have not caused any dramatic effect on the structure of wave packet.
The present results are very encouraging and call for including the
effect of the pulse on the nuclear wave packet in determining a more
refined chirp function useful in particular for longer pulses and
also for high intensity pulses. This will include new studies using
laser pulses in which the time dependence is designed to follow the
dynamical evolution of the realistic wave packet. Namely, by means
of the feedback-optimized frequency chirped laser pulses one could
try to guide the time evolution of a molecular system together with
the LICI towards a desired target state. In any case, the LICI opens
the door to investigate light-induced nonadiabatic phenomena. 
\begin{acknowledgments}
The authors acknowledge the financial support by the Deutsche Forschungsgemeinschaft
(Project ID CE10/50-3). The ELI-ALPS project (GOP-1.1.1-12/B-2012-000,
GINOP-2.3.6-15-2015-00001) is supported by the European Union and
co-financed by the European Regional Development Fund. For this work,
the supercomputing service of NIIF has been used. The authors thank
H.-D. Meyer for very helpful discussions about the MCTDH calculations.\end{acknowledgments}


\begin{thebibliography}{10}
\bibitem{Graham1} G. A. Worth and L. S. Cederbaum, \emph{Annu. Rev.
Phys. Chem}., 2004, \textbf{55}, 127.

\bibitem{Baer1} M. Baer, Beyond Born Oppenheimer: Electronic Non-Adiabatic
Coupling Terms and Conical Intersections 2006, Wiley, New York.

\bibitem{Nimrod1} N. Moiseyev, M. Sindelka and L. S. Cederbaum, \emph{J.
Phys. B.,} 2008, \textbf{41}, 221001.

\bibitem{Milan1} M. Sindelka, N. Moiseyev and L. S. Cederbaum, \emph{J.
Phys. B.,} 2011, \textbf{44}, 045603.

\bibitem{Gabi1} G. J. Halász, Á. Vibók, M. Sindelka, N. Moiseyev
and L. S. Cederbaum, \emph{J. Phys. B.,} 2011,\textbf{ 44}, 175102.

\bibitem{Gabi2} G. J. Halász, M. Sindelka, N. Moiseyev, L. S. Cederbaum
and Á. Vibók, \emph{J. Chem. Phys. A}., 2012, \textbf{116}, 2636.

\bibitem{Gabi3} G. J. Halász, Á. Vibók, M. Sindelka, L. S. Cederbaum
and N. Moiseyev, \emph{Chem. Phys.,} 2012, \textbf{399}, 146.

\bibitem{Gabi4} G. J. Halász, Á. Vibók, N. Moiseyev and L. S. Cederbaum,
\emph{Phys. Rev. A.,} 2013, \textbf{88}, 043413.

\bibitem{Natan1} A. Natan, M. R. Ware and P. H. Bucksbaum, Book of
Ultrafast Phenomena XIX Springer Proceedings in Physics Volume 2015,
\textbf{162}, 122.

\bibitem{Gabi5} G. J. Halász, Á. Vibók and L.S: Cederbaum, \emph{J.
Phys. Chem Lett.,} 2015, \textbf{6}, 348.

\bibitem{Lenz1} L. S. Cederbaum, Y. C. Chiang, Ph. V. Demekhin and
N. Moiseyev, \emph{Phys. Rev. Lett.}, 2011, \textbf{106}, 123001.

\bibitem{Philip1} Ph. V. Demekhin and L. S. Cederbaum, \emph{J. Chem.
Phys.,} 2013, \textbf{46}, 164008.

\bibitem{Kim} J. Kim, H. Tao, J. L. White, V. S. Petrovi, T. J. Martinez
and P. H. Bucksbaum, \emph{J. Phys. Chem. A., }2012, \textbf{116,}
2758. 

\bibitem{Ignacio1} M. E. Corrales, J. González-Vázquez, G. Balerdi,
I. R. Solá, R. de Nalda and L. Bañares, \emph{Nature Chem.,} 2014,
\textbf{6},785. 

\bibitem{Ignacio2} I. R. Solá, J. González-Vázquez, R. de Nalda and
L. Bañares: Strong field laser control of photochemistry. \emph{Phys.
Chem. Chem. Phys.,}\textbf{ 17,} 13183 (2015). 

\bibitem{Albert} A. Stolow, \emph{Nature Chem}., 2014, \textbf{6},
759.

\bibitem{Kim1} J. Kim, H. Tao, T. J. Martinez, P. H. Bucksbaum, \emph{J.
Phys. B.,} 2015, \textbf{48}, 164003.

\bibitem{Andris1} A. Csehi, G. J. Halász, L. S. Cederbaum and Á.
Vibók, \emph{J. Chem. Phys.}, 2015, \textbf{143}, 014305. 

\bibitem{Andris2} A. Csehi, G. J. Halász, L. S. Cederbaum and Á.
Vibók, \emph{J. Chem. Phys.}, 2016, \textbf{144}, 074309.

\bibitem{Cao1} J. Cao, C. J. Bardeen and K. R. Wilson, \emph{Phys.
Rev. Lett.}, 1998, 80, 1406.

\bibitem{Cao2} J. Cao, C. J. Bardeen and K. R. Wilson, \emph{J. Chem.
Phys}., 2000, 113, 1898.

\bibitem{krause} V. S. Malinovsky and J. L. Krause, Eur. \emph{Phys.
J. D., }2001, \textbf{14}, 147. 

\bibitem{datta} A. Datta, S. S. Bhattacharyya and B. Kim, \emph{Phys.
Rev. A., }2002, \textbf{65}, 043404. 

\bibitem{Marangos} E. Heesel, B. M. Garraway and J. P. Marangos,
\emph{J. Chem. Phys.,} 2006, \textbf{124}, 024320.

\bibitem{Adam1} A. Kirrander, H. H. Fielding and Ch. Jungen, \emph{J.
Chem. Phys.,} 2010, \textbf{132}, 024313.

\bibitem{Adam2} A. Kirrander, Ch. Jungen and H. H. Fielding, \emph{J.
Phys. B.,} 2008, \textbf{41}, 074022.

\bibitem{Adam3} A. Kirrander and H. H. Fielding, \emph{J. Phys. B.,}
2007, \textbf{40}, 897.

\bibitem{kosloff} G. Katz, M. A. Ratner and R. Kosloff, \emph{New
J. Phys. }2010, \textbf{12}, 015003.

\bibitem{kosloff1} J. Vala and R. Kosloff, \emph{Optics Express},
2001, \textbf{8}, 238.

\bibitem{kosloff2} L. Rybak, Z. Amitay, S. Amaran, R. Kosloff, M.
Tomza, R. Moszynski and C. P. Koch, \emph{Faraday Discuss. Chem. Soc.}
2011, \textbf{153}, 383. 

\bibitem{kosloff3} L. Rybak, S. Amaran, L. Levin, M. Tomza, R. Moszynski,
Z. R. Kosloff, C. P. Koch and Z. \emph{Amitay, Phys. Rev. Lett.},
2011, \textbf{107}, 273001. 

\bibitem{kosloff4} L. Levin, W. Skomorowski, L. Rybak, R. Kosloff,
C. P. Koch and Z. Amitay, \emph{Phys. Rev. Lett.}, 2015, \textbf{114},
233003.

\bibitem{prob1} V. S. Prabhudesai, U. Lev, A. Natan, B. D. Bruner,
A. Diner, O. Heber, D. Strasser, D. Schwalm, I. Ben-Itzhak, J. J.
Hua, B. D. Esry, Y. Silberberg, D. Zajfman, \emph{Phys. Rev. A.} 2010,
\textbf{81}, 023401.

\bibitem{natan} A. Natan, U. Lev, V. S. Prabhudesai, B. D. Bruner,
D. Strasser, D. Schwalm, I. Ben-Itzhak, O. Heber, D. Zajfman and Y.
Silberberg, \emph{Phys. Rev. A} 2012, \textbf{86}, 043418.

\bibitem{prob2} V. S. Prabhudesai, A. Natan, B. D. Bruner, Y. Silberberg,
U. Lev, O. Heber, D. Strasser, D. Schwalm, D. Zajfman, I. Ben-Itzhak,
\emph{J. Kor. Phys. Soc.} 2011, \textbf{59}, 2890.

\bibitem{chang} B. Y. Chang, S. Shin, J. Santamaria and I. R. Sola,\emph{
J. Phys. Chem. A} 2012, \textbf{116}, 2691. 

\bibitem{forre} S. Askeland and M. Førre, \emph{Phys. Rev. A} 2013,
\textbf{88}, 043411. 

\bibitem{zhang} C. P. Zhang, X. Y. Miao, \emph{Spect. Lett.} 2014,
\textbf{47}, 267.

\bibitem{Natan } U. Lev, L. Graham, C. B. Madsen, I. Ben-Itzhak,
B. D. Bruner. B. D. Esry, H. Frosting, O. Heber, A. Natan and V. S.
Prabhudesai, \emph{J. Phys. B} 2015, \textbf{48}, 201001.

\bibitem{Noviitsky} D. V. Novitsky, \emph{Optic Comm.} 2016, \textbf{358},
202.

\bibitem{Henriksen1} C. C. Shu and N. E. Henriksen, \emph{J. Chem.
Phys.}, 2011, \textbf{134}, 164308.

\bibitem{Henriksen2} C. C. Shu and N. E. Henriksen, \emph{J. Chem.
Phys.}, 2012, \textbf{136}, 044303.

\bibitem{Henriksen3} A. K. Tiwari , D. Dey and N. E. Henriksen, \emph{Phys.
Rev. A} 2014, \textbf{89}, 023417. 

\bibitem{Floquet} S. I. Chu, \emph{J. Chem. Phys.} 1981, \textbf{75},
2215.

\bibitem{tdm} S. I. Chu, C. Laughlin, and K. Datta,\emph{ Chem. Phys.
Lett.} 1983, \textbf{98}, 476.

\bibitem{pot} F. V. Bunkin and I. I. Tugov, \emph{Phys. Rev. A} 1973,
\textbf{8}, 601.

\bibitem{Dieter1} H. D. Meyer, U. Manthe and L. S. Cederbaum, Chem.
Phys. Lett. 1990\textbf{, 165}, 73. 

\bibitem{Dieter2} U. Manthe, H. D. Meyer and L. S. Cederbaum, J.
Chem. Phys. 1992, \textbf{97}, 3199. 

\bibitem{Dieter3} M. H. Beck, A. Jäckle, G. A. Worth and H. D. Meyer,
Phys. Rep. 2000, \textbf{324}, 1.

\bibitem{Dieter4} G. A. Worth et al. The MCTDH package, version 8.2;
University of Heidelberg: Heidelberg, Germany, 2000. H. D. Meyer et
al. The MCTDH package, versions 8.3 and 8.4; University of Heidelberg,
Germany, 2002 and 2007. http://mctdh.uni-hd.de/. 

\bibitem{Dieter5} Multidimensional Quantum Dynamics: MCTDH Theory
and Applications, edited by H. D. Meyer, F. Gatti, G. A. Worth, (Wiley-VCH:
Weinheim, 2009).

\bibitem{Schinke} R. Schinke, Photodissociation Dynamics, Cambridge
University Press, Cambridge, 1993.

\bibitem{Gabriel} G. G. Balint-Kurti, R. N. Dixon and C. C. Marston,\emph{
J. Chem. Soc., Faraday Trans.} 1990, \textbf{86}, 1741.

\bibitem{Gatti} O. Vendrell, F. Gatti and H.-D. Meyer, \emph{J. Chem.
Phys}. 2007, \textbf{127}, 184303. 

\bibitem{Pulse} D. B. Milosevic, G. G. Paulus, D. Bauer, and W. Becker,
\emph{J. Phys. B} 2006,\textbf{ 39}, R203.\end{thebibliography}
\end{document}